\documentclass[conference]{IEEEtran}
\IEEEoverridecommandlockouts
\usepackage{fancyhdr}
\usepackage{cite}
\usepackage{amsmath,amssymb,amsfonts}
\usepackage{algorithmic}
\usepackage{graphicx}
\usepackage{textcomp}
\usepackage{xcolor}
\usepackage{eurosym}
\usepackage{adjustbox}
\usepackage{array}    
\usepackage{fancyhdr}
\usepackage{multirow}
\usepackage{hyperref}
\usepackage{tikz}
\usepackage{array}
\usepackage[caption=false,font=footnotesize]{subfig}
\usepackage{stfloats}
\usepackage{url}
\newcommand{\blueref}[1]{\textcolor{black}{(}\ref{#1}{)}}
\newcommand{\figref}[1]{Fig.~\ref{#1}}
\newcommand{\tabref}[1]{Table~\ref{#1}}
\usetikzlibrary{decorations.pathreplacing} 
\usetikzlibrary{shapes.geometric} 
\usetikzlibrary{decorations.pathreplacing, shapes.geometric} 

\def\BibTeX{{\rm B\kern-.05em{\sc i\kern-.025em b}\kern-.08em
    T\kern-.1667em\lower.7ex\hbox{E}\kern-.125emX}}

\usepackage{titlesec}

\usepackage{amsmath}
\renewcommand{\thesection}{\arabic{section}}
\renewcommand{\thesubsection}{\arabic{section}.\arabic{subsection}}
\renewcommand{\thesubsubsection}{\arabic{section}.\arabic{subsection}.\arabic{subsubsection}}

\titleformat{\section}[block]{\bfseries\normalsize}{\thesection\quad}{0pt}{}  
\titleformat{\subsection}[block]{\itshape\normalsize}{\thesubsection\quad}{0pt}{}  
\titleformat{\subsubsection}[block]{\itshape\normalsize}{\thesubsubsection\quad}{0pt}{} 

\begin{document}
\rmfamily

\title{{\fontsize{16}{19}\selectfont \textbf{Betting vs. Trading: Learning a Linear Decision Policy \\ for Selling Wind Power and Hydrogen}}}

\author{\IEEEauthorblockN{Yannick Heiser,
Farzaneh Pourahmadi, and
Jalal Kazempour}
\IEEEauthorblockA{Technical University of Denmark, Kgs. Lyngby, Denmark\\
{\{yahei, farpour, jalal\}@dtu.dk}}}

\maketitle
\thispagestyle{fancy}

\begin{abstract}
We develop a bidding strategy for a hybrid power plant combining co-located wind turbines and an electrolyzer, constructing a price-quantity bidding curve for the day-ahead electricity market while optimally scheduling hydrogen production. Without risk management, single imbalance pricing leads to an all-or-nothing trading strategy, which we term “betting". To address this, we propose a data-driven, pragmatic approach that leverages contextual information to train linear decision policies for both power bidding and hydrogen scheduling. By introducing explicit risk constraints to limit imbalances, we move from the all-or-nothing approach to a “trading" strategy, where the plant diversifies its power trading decisions. We evaluate the model under three scenarios: when the plant is either conditionally allowed, always allowed, or not allowed to buy power from the grid, which impacts the green certification of the hydrogen produced. Comparing our data-driven strategy with an oracle model that has perfect foresight, we show that the risk-constrained, data-driven approach delivers satisfactory performance.
\end{abstract}


\vspace{1mm}
\section{Introduction}

The optimal bidding decision-making problem for renewable energy sources in the day-ahead electricity market under \textit{dual} imbalance pricing is often framed as a “newsvendor" problem \cite{Pinson2007}. In this context, any production imbalance, i.e., a deviation from the day-ahead schedule in real time, incurs a penalty. This creates a strong incentive to bid as closely as possible to the forecasted power production, despite the inherent uncertainty in these forecasts at the day-ahead stage. To effectively model and manage this uncertainty in the newsvendor framework, a variety of trading strategies have been developed. These include approaches such as probabilistic forecasting \cite{Pinson2007}, two-stage stochastic optimization \cite{Morales2010}, robust optimization \cite{Sun2021}, distributionally robust optimization \cite{Pinson2023}, linear decision policies \cite{Muñoz2020}, and online learning \cite{cms}.


However, most European countries have transitioned to \textit{single} imbalance pricing, in line with the European Balancing Guideline 2017/2195 \cite{european_balancing_guideline_2017}. In Denmark, for example, the imbalance pricing system was initially dual but switched to single pricing in November 2021. Single imbalance pricing introduces arbitrage opportunities between the day-ahead and balancing markets, enabling market participants to profit by forecasting which market will have a higher price. However, this strategy requires accurately predicting the direction of grid imbalance---whether the system operator will face a power surplus or deficit in real time---a task that is complex, if not impossible. As a result, the power trading problem under single imbalance pricing for a trader seeking to exploit these arbitrage opportunities becomes a “betting" problem, where the producer commits to selling \textit{all-or-nothing} of its power in the day-ahead market based on the expectation that the day-ahead price will be either higher or lower than the balancing market price in the same hour. This high level of risk highlights the need for methods to mitigate exposure in energy trading, raising the critical question: how can a power producer avoid falling into betting behavior when trading in the electricity market?


Classical risk-aware optimization problems typically incorporate a risk measure, such as conditional value-at-risk (CVaR) \cite{cvar}, into the objective function to mitigate risk. However, this approach often retains an all-or-nothing trading strategy, merely adjusting the critical threshold for the expected price difference at which a trader shifts its entire position from one market to another. While this reduces risk exposure, it does not necessarily  eliminate the binary nature of decision-making. For instance, \cite{Browell_2018} develops risk-constrained strategies that mitigate risk by adjusting outputs, yet they still result in binary trading outcomes and require detailed information on both day-ahead and balancing prices, as well as the system state (long or short).

In contrast, we propose an approach that explicitly enforces \textit{risk constraints} to limit the feasible imbalance settlement space, thereby reducing exposure. To the best of our knowledge, this is the first work to apply risk constraints in a way that eliminates the all-or-nothing “betting" strategy under single imbalance pricing, transforming it into a “trading" strategy. This enables market participants to \textit{diversify} their power trading decisions, crucially removing the binary nature of trading decisions typically observed under single imbalance pricing.


As a case study, we consider the optimal bidding strategy decision-making problem for a hybrid power plant (HPP) that combines co-located wind turbines and an electrolyzer. The HPP operator aims to maximize profit by making optimal bidding decisions in electricity markets, either selling (or, if allowed, buying) power, while also selling hydrogen through bilateral contracts with hydrogen off-takers. We examine three operating scenarios: the HPP operator is either allowed to buy power from the grid, restricted to only selling to the grid, or subject to a conditional buying constraint. The primary sources of uncertainty in this problem are wind power generation and electricity market prices.


Building on \cite{HELGREN2024}, which develops a data-driven trading model for an HPP under dual imbalance pricing, we implement linear decision policies \cite{kuhn} that directly map available contextual information to the trading decision variables. While conventional decision-making methods under uncertainty, such as stochastic and robust optimization, can be complex and often require probabilistic forecasts, linear decision policies offer a \textit{more practical} approach for practitioners. These policies are simple, easy to implement, computationally efficient, and transparent for operators, all while effectively addressing the complexities of uncertainty. Additionally, linear decision policies are highly scalable making them suitable for any portfolio size. 
In contrast to \cite{HELGREN2024}, which focuses on dual imbalance pricing, our study considers single imbalance pricing, where a betting strategy typically emerges. To mitigate the risks associated with such strategies, we incorporate explicit risk constraints into our data-driven approach, transforming the all-or-nothing behavior into a more diversified trading strategy.


The rest of this paper is structured as follows: Section~\ref{sec:overall framework} provides an overview of the proposed framework. Section~\ref{Sec:Training} details the training process. Section~\ref{Sec:Test+Feasibility} outlines the testing procedure and how a feasible solution is restored, if necessary. Section~\ref{sec:results} presents the numerical results, and finally, Section~\ref{sec:conclusion} concludes the paper.

\textit{Note}: In the following, symbols with a hat $\hat{(.)}$ denote realizations, while those with a tilde $\tilde{(.)}$ represent forecasts. Bold symbols denote vectors, symbols with a breve $\Breve{(.)}$ define subsets of vectors, $(.)^\top$ denotes the transpose operator, $|.|$ represents the cardinality of a set, lower-case symbols denote variables, and upper-case or Greek symbols represent input data.

\vspace{1mm}
\section{Overall framework}\label{sec:overall framework}


The HPP consists of co-located assets, including wind turbines, which are coupled behind-the-meter with an alkaline electrolyzer. The hydrogen produced by the HPP is sold at a fixed price under a bilateral agreement. This is the typical case of hydrogen trading with no hydrogen market being established yet. We enforce a minimum daily hydrogen production requirement as part of the bilateral contract, which, for example, could be determined by the daily loading of tube trailers used to transport the stored hydrogen to customers.  For simplicity, we assume that all the produced hydrogen is stored in industrial gas bottles at around 200 bars corresponding to a constant loss factor.  The hydrogen production curve of the electrolyzer, which shows the amount of hydrogen produced as a function of consumed power, is inherently non-convex. However, we model it using a linear approximation, represented by a set of linear cuts (an outer approximation). This linear relaxation method is provably exact for common operating conditions \cite{RAHELI2023}. For simplicity, we assume that the electrolyzer is always operational, meaning there is a non-zero minimum power consumption and no ramping constraints.


In the day-ahead stage, the HPP operator must make decisions on electricity market bids and schedule hydrogen production with the goal of maximizing total profit from the sale of both electricity and hydrogen. The operator can sell wind power in the day-ahead market, while the option to buy power from the grid is subject to three possible operating conditions:
\begin{enumerate} 
\item \textbf{Buying permitted:} Buying power from the grid is allowed, with no restrictions. 
\item \textbf{Buying not permitted:} Buying power from the grid is not allowed, meaning the decision-making process is \textit{restricted} to selling only (referred to hereafter as \textit{res}). 
\item \textbf{Buying conditionally permitted:} Buying power from the grid is \textit{conditionally} allowed, under specific circumstances (referred to hereafter as \textit{cond}). This condition aligns with the ``E.U. policies for Renewable Hydrogen" published in 2023 \cite{eu_rules_renewable_hydrogen_2023}. 
\end{enumerate}
While the hydrogen produced under the second and third operating conditions can be labeled ``green", this is not necessarily the case in the first operating condition.


The HPP operator must make electricity market bidding decisions for all hours of day $\rm{D}$ on the day before, typically by noon on $\rm{D}$-1 in the Nord Pool market. However, at the time these decisions are made, the actual wind power production for day $\rm{D}$ remains uncertain. Any deviations from the day-ahead power quantity schedule must be settled in the balancing market, which follows a single imbalance pricing mechanism, as is commonly applied in most European countries today.\


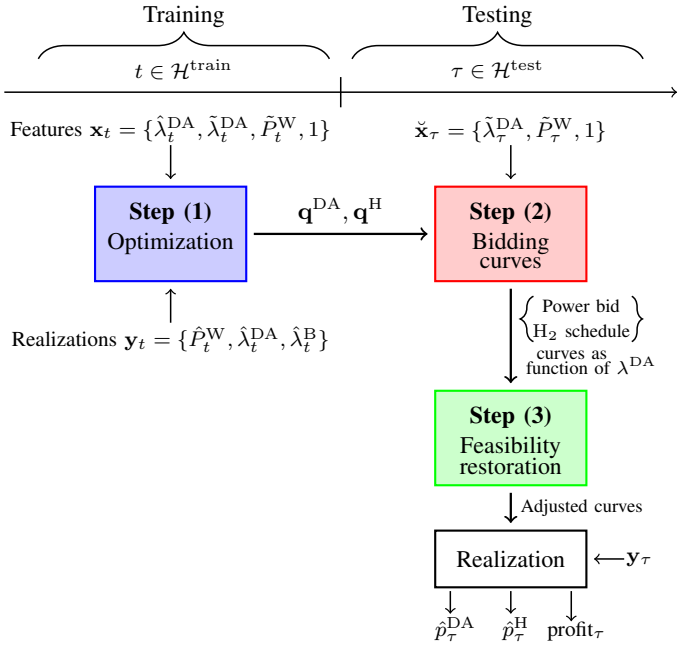
\begin{figure}[!t]
\centering
\begin{tikzpicture}
\filldraw[fill=blue!20, draw=blue, thick] (-0.5,-0.5) rectangle (1.5,0.75);
\filldraw[fill=red!20, draw=red, thick] (4,-0.5) rectangle (6,0.75);
\filldraw[fill=green!20, draw=green, thick] (4,-3.2) rectangle (6,-1.95);

\draw[thick] (4,-3.8) rectangle (6,-4.55);
\draw[->,line width=0.2mm] (-1.7,2) -- (7.2,2);
\draw[line width=0.2mm] (2.75,2.2) -- (2.75,1.8);
\draw [decorate,decoration={brace,amplitude=10pt,raise=4pt},line width=0.2mm] (-1.25,2.3) -- (2.6,2.3) node[midway,yshift=20pt]{\small Training} node[midway,yshift=0]{\footnotesize $t \in \mathcal{H}^{\rm{train}}$};
\draw [decorate,decoration={brace,amplitude=10pt,raise=4pt},line width=0.2mm] (2.9,2.3) -- (6.75,2.3) node[midway,yshift=20]{\small Testing} node[midway,yshift=0]{\footnotesize $\tau \in \mathcal{H}^{\rm{test}}$};
\node at (0.5, 0) {\small Optimization};
\node at (0.5, 0.4) {\small\textbf{Step (1)}};
\node at (5, 0.4) {\small \textbf{Step (2)}};
\node at (5, 0) {\small Bidding};
\node at (5, -0.25) {\small curves};
\node at (5, -2.7) {\small Feasibility};
\node at (5, -2.95) {\small restoration};
\node at (5, -2.3) {\small \textbf{Step (3)}};

\node at (5, -4.175) {\small Realization};
\draw[->, line width=0.3mm, scale=1] (1.6, 0.125) -- (3.9, 0.125) node[midway, above] {\small$\mathbf{q}^{\rm{DA}}, \mathbf{q}^{\rm{H}}$};
\draw[->, line width=0.2mm, scale=1] (0.5, 1.3) -- (0.5, 0.85) node[midway,above,yshift=4]{\footnotesize Features $\mathbf{x}_t = \{\hat{\lambda}_t^{\rm{DA}},\Tilde{\lambda}_t^{\rm{DA}},\Tilde{P}_t^{\rm{W}},1\}$};
\draw[->, line width=0.2mm, scale=1] (0.5, -1.05) -- (0.5, -0.6) node[midway,below,yshift=-4]{\footnotesize Realizations $\mathbf{y}_t = \{\hat{P}_t^{\rm{W}},\hat{\lambda}_t^{\rm{DA}},\hat{\lambda}_t^{\rm{B}}\}$};
\draw[->, line width=0.2mm, scale=1] (5, 1.3) -- (5, 0.85) node[midway,above,yshift=4]{\footnotesize $\Breve{\mathbf{x}}_\tau = \{\Tilde{\lambda}_\tau^{\rm{DA}},\Tilde{P}_\tau^{\rm{W}},1\}$};
\draw[->, line width=0.3mm, scale=1] (5, -0.6) -- (5, -1.85) node[midway,yshift=-7,xshift=6,right]{\scriptsize curves as}
node[midway,yshift=-11,xshift=2,right]{\scriptsize  function of $\lambda^{\rm{DA}}$};
\draw [decorate,decoration={brace,amplitude=4pt,raise=37},line width=0.2mm] (5.3,-0.7) -- (5.3,-1.3) node[midway,yshift=5,xshift=18]{\scriptsize Power bid} node[midway,yshift=-5,xshift=18]{\scriptsize $\rm{H}_2$ schedule};
\draw [decorate,decoration={brace,amplitude=5pt,raise=0pt},line width=0.2mm] (5.3,-1.3) -- (5.3,-0.7);
\draw[->, line width=0.3mm, scale=1] (5, -3.3) -- (5, -3.7) node[midway,yshift=0,right]{\scriptsize Adjusted curves};
\draw[->, line width=0.2mm, scale=1] (6.5, -4.175) -- (6.1, -4.175) node[midway,xshift=3,right]{\footnotesize $\mathbf{y}_\tau$};
\draw[->, line width=0.2mm, scale=1] (5.8, -4.6) -- (5.8,-5) node[midway,xshift=2,yshift=-2,below]{\footnotesize profit$_\tau$};
\draw[->, line width=0.2mm, scale=1] (4.2, -4.6) -- (4.2,-4.9) node[midway,xshift=2,yshift=-2,below]{\footnotesize $\hat{p}_\tau^{\rm{DA}}$};
\draw[->, line width=0.2mm, scale=1] (5, -4.6) -- (5,-4.9) node[midway,xshift=2,yshift=-2,below]{\footnotesize $\hat{p}_\tau^{\rm{H}}$};
\end{tikzpicture}
\caption{The overall framework of the proposed model:  The feature vector $\mathbf{x}_t$  and the uncertainty realization vector $\mathbf{y}_t$ during the training period $t \in \mathcal{H}^{\rm{train}}$ are used to determine the optimal linear policies for power trading ($\mathbf{q}^{\rm{DA}}$) and hydrogen scheduling ($\mathbf{q}^{\rm{H}}$). In the testing period
$\tau \in \mathcal{H}^{\rm{test}}$,
a subset of the feature vector,  $\Breve{\mathbf{x}}_\tau$,  is used to construct a bidding curve (power trade as a function of day-ahead electricity price)  to be submitted to the day-ahead market, as well as to determine the hydrogen production schedule. If these outcomes are infeasible during testing, a feasible solution is restored. Once the realizations $\mathbf{y}_\tau$ of wind power production, day-ahead prices,  and imbalance prices are available during the testing phase, the accepted power bid $\hat{p}_\tau^{\rm{DA}}$, the hydrogen schedule $\hat{p}_\tau^{\rm{H}}$ and the final profit are calculated.}
\vspace{-2mm}
\label{fig:framework}
\end{figure}

A pragmatic way to manage uncertainty is to use (linear) decision policies, where the HPP operator learns from historical data how to form an optimal policy that connects available contextual information at the time of decision-making to the corresponding bidding decisions. This information builds a set of features, which can serve as input for training a machine learning model to make predictions. Hereafter, we will use the terms ``features" and ``contextual information", interchangeably. These features can include any data that exhibits spatial or temporal linear correlation with the decision variables, such as aggregated wind power forecasts and electricity price forecasts. We refer the interested reader to \cite{overview} for a comprehensive overview of contextual optimization methods for decision-making under uncertainty.

We develop a data-driven model for the HPP operator that leverages available contextual information. Our proposed framework, as illustrated in  \figref{fig:framework}, consists of three main steps: (1) \textbf{model training} through optimization using the training data during the training period $t \in \mathcal{H}^{\rm{train}}$, which includes the feature vector $\mathbf{x}_t$ and the uncertainty realization vector $\mathbf{y}_t$.  The feature vector includes forecasted prices, $\Tilde{\lambda}_t^{\rm{DA}}$, and wind power production data, $\Tilde{P}_t^{\rm{W}}$, as well as the realized day-ahead prices $\hat{\lambda}_t^{\rm{DA}}$ and a constant 1 as described in Section \ref{sec: feature set}. The realization vector includes the realized prices of the day-ahead market,  $\hat{\lambda}_t^{\rm{DA}}$, and balancing market, $\hat{\lambda}_t^{\rm{DA}}$, as well as the realized power production, $\hat{P}_t^{\rm{W}}$.  This first step determines the optimal linear decision policies for power trading $\mathbf{q}^{\rm{DA}}$ and hydrogen scheduling $\mathbf{q}^{\rm{H}}$; (2) \textbf{model testing} during the testing period $\tau \in \mathcal{H}^{\rm{test}}$, which constructs a bidding curve based on the given  policies and a subset of the feature vector $\mathbf{\breve{x}}_\tau$  excluding the realized day-ahead price as described in Section \ref{sec:bidding curve} ; and finally, (3) \textbf{feasibility restoration} of the model output if necessary.  As explained in Section \ref{sec: Feasibility Restoration Step}, the model testing output of the previous step can form an infeasible bidding curve, which requires a feasibility restoration step.  Using the uncertainty realization $\mathbf{y}_\tau$, the final accepted power bid $\hat{p}_t^{\rm{DA}}$, hydrogen production schedule $\hat{p}_t^{\rm{H}}$ and profit can be calculated.
It is worth mentioning that both the training and testing steps rely on available historical data, so the testing phase can be seen as a form of back-testing. The following sections describe the concepts behind each step, beginning with Step (1) in Section~\ref{Sec:Training}, and Steps (2) and (3) in Section~\ref{Sec:Test+Feasibility}.
\vspace{1mm}
\section{Training Step}\label{Sec:Training}
\subsection{Feature Set}\label{sec: feature set}
Linear decision policies can be viewed as a form of linear regressors used to fit the optimal decision in an $N$-dimensional feature space. This is achieved by training the model on historical data, which includes both the features and the uncertain parameters.
We first define the feature vector for hour $t \in \mathcal{H}^{\rm{train}}$ as:
\begin{equation}\label{eq:1}
    \mathbf{x}_t = \big\{x_{t,1},...,x_{t,N\!-\!1},1\big\} \in \mathbb{R}^{N},
\end{equation}
consisting of $N$-1  features, as well as a constant feature $1$ added as the $N^{\rm{th}}$ element, $x_{t,N}$, to account for the intercept. As illustrated in \figref{fig:framework}, our feature vector for hour $t$ includes the \textit{realized} day-ahead electricity price $\hat{\lambda}_t^{\rm{DA}}$, the \textit{forecasted} day-ahead price $\Tilde{\lambda}_t^{\rm{DA}}$, the \textit{forecasted} wind power, $\Tilde{P}_t^{\rm{W}}$, and the constant feature $1$. For the forecast-based features, we can include various additional forecasts, such as different price and wind forecasts from multiple vendors, as well as the local wind power forecast for the HPP and the aggregated regional and national wind power forecasts. Note that the realized day-ahead price is intentionally included in the feature vector to construct a bidding curve during the testing phase. 
\vspace{1mm}
\subsection{Linear Decision Policies}
Let $p_t^{\rm{DA}}, p_t^{\rm{H}}  \in \mathbb{R}$, both in MW, represent the day-ahead power trade and the power consumption of the electrolyzer in hour $t$, respectively.
Given the feature vector in \blueref{eq:1}, we define the linear decision policies by:
\begin{subequations}\label{eq:2}
\begin{align}
    p_t^{\rm{DA}} = \mathbf{q}_{j,k}^{\rm{DA}} \ \mathbf{x}_t^\top \ \ & \ \ \ \forall t \in \mathcal{H}^{\rm{train}}\label{eq:2a}\\
    p_t^{\rm{H}} = \mathbf{q}_{j,k}^{\rm{H}} \ \mathbf{x}_t^\top \ \ & \ \ \ \forall t \in \mathcal{H}^{\rm{train}},\label{eq:2b}\
\end{align}
\end{subequations}
where  $k \!\in\!\mathcal{K}$ represents the index for the power price domains. For example, consider two predefined power price domains: (0,100) and (100,1000) €/MWh, with $\mathcal{|K|}=2$. The price thresholds for these domains can be viewed as hyperparameters that must be carefully tuned.  Note that $t$
is the index for all hours in the historical training data, which may span multiple days or even months, e.g., two months, where $t \in \{1,...,1440\}$, while $j\!=\!(t\!-\!1)\!\mod{24}\!+\!1\in\!\{1,...,24\}$ represents the hours within a single day.  Accordingly, we define the power trading policy vector $\mathbf{q}_{j,k}^{\rm{DA}}  \in \mathbb{R}^{N}$ and the hydrogen scheduling policy vector $\mathbf{q}_{j,k}^{\rm{H}}  \in \mathbb{R}^{N}$ for each hour $j$ and price domain $k$. For example, consider a historical data point for $t\!=\!25$,  where the realized day-ahead price is €150/MWh. This data point contributes to learning policies for $j\!=\!1$ (the first hour of the day) and $k\!=\!2$ as the realized price falls into the second price domain of (100,1000). By defining such policies for each hour $j$ and price domain $k$, we introduce an additional degree of freedom into the learning model. Moreover, as mentioned earlier, obtaining policy values across price domains enables the construction of a price-quantity bidding curve. The training optimization problem to determine optimal policies, corresponding to Step (1) in \figref{fig:framework}, is provided in the following subsection.



\vspace{1mm}
\subsection{Betting}
We begin with the betting model without any risk constraints, where the HPP operator makes all-or-nothing trading decisions in the day-ahead stage.  The operator aims to maximize profit from trading power $p_t^{\rm{DA}}$ in the day-ahead market at price $\hat{\lambda}_t^{\rm{DA}} \in \mathbb{R}_+$, by selling hydrogen $h_t  \in \mathbb{R}_+$, in kg, at a fixed price $\lambda^{\rm{H}} \in \mathbb{R}_+$, in  €/kg, and settling power deviation $\Delta p_t\in \mathbb{R}$ in the balancing market at the single imbalance price $\hat{\lambda}_t^{\rm{B}}\in \mathbb{R}_+$. Note that $p_t^{\rm{DA}}$ and $\Delta p_t$ are positive when selling power, but take negative values when buying power from the grid, if allowed. The resulting linear program is:
%
%
%
\begingroup
\allowdisplaybreaks
\begin{subequations}\label{eq:3}
\begin{align}
    &\underset{\Omega} {\rm{Maximize}} \ \underset{t\in\mathcal{H}^{\rm{train}}}{\sum} p_t^{\rm{DA}} \hat{\lambda}_t^{\rm{DA}} + h_t \lambda^{\rm{H}} +  \Delta p_t \hat{\lambda}_t^{\rm{B}} \label{eq:3a}
\end{align}
\begin{align}
    \text{s.t.} \ \ \ \  & \text{\blueref{eq:2a}-\blueref{eq:2b}} & \label{eq:3b} \\
    & \hat{P}_t^{\rm{W}} \geq p_t^{\rm{DA}} + \Delta p_t + p_t^{\rm{H}} & \forall t \in \mathcal{H}^{\rm{train}}\label{eq:3c}\\
    & p_t^{\rm{DA}} \leq \overline{P}^{\rm{W}} & \forall t \in \mathcal{H}^{\rm{train}}\label{eq:3d}\\
    & \underline{P}^{\rm{H}} \leq p_t^{\rm{H}} \leq \overline{P}^{\rm{H}}  & \forall t \in \mathcal{H}^{\rm{train}}\label{eq:3e}\\
    & h_t \leq A_s p_t^{\rm{H}} + B_s & \forall s \in \mathcal{S}, \ \forall t \in \mathcal{H}^{\rm{train}}\label{eq:3f}\\
    & \sum_{t=24d}^{24(d+1)} h_t \geq \underline{D} & \forall d \in \mathcal{D}^{\rm{train}},\label{eq:3g}
\end{align}
\end{subequations}
\endgroup
where  $\Omega = \{p_t^{\rm{DA}},\Delta p_t,h_t,p_t^{\rm{H}},\mathbf{q}_{j,k}^{\rm{DA}},\mathbf{q}_{j,k}^{\rm{H}}\}$ is the set of decision variables. The objective function The objective function \blueref{eq:3a} incorporates three revenue terms from the day-ahead electricity market, hydrogen sales, and the balancing market. 
Constraint \blueref{eq:3b} defines the structure of the linear decision policies, as explained earlier. 
Constraint \blueref{eq:3c} enforces the power balance of the HPP, where the parameter $\hat{P}_t^{\rm{W}}$ represents the realization of wind power. This value must be greater than or equal to the power traded in the day-ahead and balancing markets, as well as the power consumed by the electrolyzer. This inequality allows for the curtailment of wind power if necessary. Constraint \blueref{eq:3d} sets the upper bound for the traded day-ahead power $p_t^{\rm{DA}}$ to be equal to the installed wind capacity, $\overline{P}^{\rm{W}}$.
Constraint \blueref{eq:3e} defines the operating range for the electrolyzer's power consumption, $p_t^{\rm{H}}$, to lie between $\underline{P}^{\rm{H}}$ and $\overline{P}^{\rm{H}}$. Constraint  \blueref{eq:3f} enforces a set of linear cuts, indexed by $s\!\in\!\mathcal{S}$, where the parameters $A_s$ and $B_s$ represent the slope and intercept of cut $s$, respectively. These cuts linearize the hydrogen production curve of the electrolyzer, which is originally non-convex. For further details on this linear relaxation method, see \cite{RAHELI2023}. Finally, \blueref{eq:3g} enforces the minimum hydrogen production requirement, $\underline{D}$, for each day $d \in \mathcal{D}^{\rm{train}}$ within the training period. 
%
%

We now introduce additional constraints to \blueref{eq:3}, based on the underlying operating conditions related to buying power from the grid. As noted in Section \ref{sec:overall framework}, we consider three possible conditions: buying permitted, buying not permitted (\textit{res}), and buying conditionally permitted (\textit{cond}).

\begin{enumerate}
\item  \textbf{Buying permitted ($\mathcal{B}$):}\
We formulate the training model for betting under an operating condition in which buying power from the grid is permitted. For this, we add the following constraint:
\begin{align}\label{eq:5}
        & p_t^{\rm{DA}} \geq -\overline{P}^{\rm{H}},  & \forall t \in \mathcal{H}^{\rm{train}}.
\end{align}
This constraint specifies that the lower bound of $p_t^{\rm{DA}}$ is equal to the negative of the electrolyzer's capacity. In other words, the maximum power that can be bought from the grid is constrained by the electrolyzer's maximum consumption level. Note that $\Delta p_t$ can also be negative, but it is not necessarily bounded by $\overline{P}^{\rm{H}}$, as this is already constrained by  (\ref{eq:3c}). Therefore, no additional constraint is required for $\Delta p_t$. The resulting model, i.e., the optimization problem \blueref{eq:3} with the additional constraint \blueref{eq:5}, is referred to as the betting model $\mathcal{B}$, where buying power from the grid is allowed. 

\item \textbf{Buying not permitted ($\mathcal{B}^{\rm{res}}$):}\
To prevent buying power from the grid in both the day-ahead and balancing stages, we introduce the following two constraints:
\begin{subequations}\label{eq:6}
    \begin{align}
    & p_t^{\rm{DA}} \geq 0,  & \forall t \in \mathcal{H}^{\rm{train}},\label{eq:6a}\\
    & \Delta p_t \geq 0, &  \forall t \in \mathcal{H}^{\rm{train}}.\label{eq:6b}
\end{align}
\end{subequations}
The resulting betting model, $\mathcal{B}^{\rm{res}}$, includes \blueref{eq:3} with two additional constraints, as specified in \blueref{eq:6}.

%

\item \textbf{Buying conditionally permitted ($\mathcal{B}^{\rm{cond}}$):}
According to the ``E.U. policies for Renewable Hydrogen" \cite{eu_rules_renewable_hydrogen_2023} published by the European Parliamentary Research Service in 2023, the hydrogen produced can still be labeled as green when buying power from the grid, provided the electricity price is lower than or equal to $\lambda^{\rm{S}}=\text{€}20$/MWh.
To model this price condition, we introduce the auxiliary binary variable $b_t \in [0,1]$ and add it to the variable set  $\Omega$, which indicates when this price condition holds. We then add the following constraints with Big-\textit{M}s:
\begin{subequations}\label{eq:7}
\begin{align}
    & \frac{\lambda^{\rm{S}}\!-\!\lambda_t^{\rm{DA}}}{M} \!\leq\! b_t \!\!\leq \! \frac{\lambda^{\rm{S}}-\lambda_t^{\rm{DA}}\! + \!M \!- \!\varepsilon}{M},\!\!\! \! &  \forall t \in \mathcal{H}^{\rm{train}},\label{eq:7a}\\
    & \Delta p_t \geq (-\overline{P}^{\rm{W}}-\overline{P}^{\rm{H}}) b_t, &  \forall t \in \mathcal{H}^{\rm{train}},\label{eq:7b}\\
    & p_t^{\rm{DA}} \geq -\overline{P}^{\rm{H}}b_t, &  \forall t \in \mathcal{H}^{\rm{train}}.\label{eq:7c}
\end{align}
\end{subequations}
Constraints \blueref{eq:7} represent the condition that the green electricity grid requirement is met when $b_t = 1$, while $b_t = 0$ indicates the opposite. The parameter $M$ is chosen to be the maximum of the realized day-ahead price, i.e.,  $M = \max(\lambda_t^{\rm{DA}}| t \in \mathcal{H}^{\rm{train}})$, and $\varepsilon>0$ is a very small constant. Adding constraints \blueref{eq:7} to the optimization model \blueref{eq:3} results in the betting model $\mathcal{B}^{\rm{cond}}$.  
Note that the introduction of the binary variable transforms the resulting model $\mathcal{B}^{\rm{cond}}$ into a mixed-integer linear program, while the models $\mathcal{B}$ and $\mathcal{B}^{\rm{res}}$ remain linear programs.
\end{enumerate}

It is important to note that we introduce two additional non-negative slack variables in the models $\mathcal{B}^{\rm{res}}$  and $\mathcal{B}^{\rm{cond}}$ to ensure feasibility, as power consumption from the grid is restricted. The first slack variable is added to the constraint enforcing the minimum power consumption of the electrolyzer, as specified in \blueref{eq:3e}. This variable takes a non-zero value when there is insufficient wind power production to meet the electrolyzer's minimum operating power requirement. The second slack variable is introduced to the constraint enforcing the minimum daily hydrogen production requirement, as specified in \blueref{eq:3g}. This variable  takes a non-zero value on days with insufficient wind power to meet the production target. These two slack variables are penalized in the objective function \blueref{eq:3a} by multiplying them with large constants, ensuring that their optimal values remain zero unless absolutely necessary to maintain feasibility.

\vspace{1mm}
\subsection{Trading}\label{sec:trading}
To avoid a betting strategy characterized by all-or-nothing bidding decisions in the day-ahead stage, we enforce three distinct sets of constraints to limit the magnitude of the power imbalance to be settled in the balancing market. We refer to these as \textit{risk constraints}, as they help transition from the betting models to trading models by diversifying trading decisions in the day-ahead stage. These risk constraints limit either the expected value, CVaR, or the maximum amount of power imbalance (either short or long) in the balancing market.

We first introduce the new auxiliary variable $\Delta p_t^{\rm{ABS}} \in \mathbb{R}$ to be added to the variable set $\Omega$. This variable represents the absolute value of the imbalance in the balancing market and is defined by the following linear constraints: 
\begin{subequations}\label{eq:8}
\begin{align}
    & \Delta p_t^{\rm{ABS}} \geq \Delta p_t, &  \forall t \in \mathcal{H}^{\rm{train}},\label{eq:8a}\\
    &\Delta p_t^{\rm{ABS}} \geq -\Delta p_t, &  \forall t \in \mathcal{H}^{\rm{train}}.\label{eq:8b}
\end{align}
\end{subequations}
With this, the three different risk constraints can be formulated as follows:
\begin{enumerate}
    \item \textbf{Mean imbalance ($\mathcal{T}_{\rm{mean}}$, $\mathcal{T}_{\rm{mean}}^{\rm{res}}$, $\mathcal{T}_{\rm{mean}}^{\rm{cond}}$):}\
    Here, we aim at limiting the mean of $\Delta p_t^{\rm{ABS}}$ such that:
\begin{align}\label{eq:9}
    & \frac{\sum_{t\in\mathcal{H}^{\rm{train}}} \Delta p_t^{\rm{ABS}}}{|\mathcal{H}^{\rm{train}}|} \leq \overline{\Delta P}^{\rm{mean}},
\end{align}
where the parameter $\overline{\Delta P}^{\rm{mean}}$ represents the given upper limit for the mean of $\Delta p_t^{\rm{ABS}}$. The model that incorporates the risk constraint \blueref{eq:9} into the optimization problem \blueref{eq:3} is referred to as the trading model with the mean constraint. Depending on the specific operational condition related to the permissibility of buying power from the grid, this leads to three possible trading models: 
$\mathcal{T}_{\rm{mean}}$ when buying is allowed, $\mathcal{T}_{\rm{mean}}^{\rm{res}}$ when buying is not allowed, and $\mathcal{T}_{\rm{mean}}^{\rm{cond}}$ when buying is conditionally allowed.
\\

\item \textbf{CVaR ($\mathcal{T}_{\rm{CVaR}}$, $\mathcal{T}_{\rm{CVaR}}^{\rm{res}}$, $\mathcal{T}_{\rm{CVaR}}^{\rm{cond}}$):}\
Here, we limit the CVaR of $\Delta p_t^{\rm{ABS}}$ such that:
\begin{subequations}\label{eq:10}
\begin{align}
    & VaR \ + \frac{1}{(1-\alpha)|\mathcal{H}|} \sum_{t\in\mathcal{H}^{\rm{train}}} \xi_t \leq \overline{\Delta P}^{\rm{CVaR}},\label{eq:10a}
\end{align}
\begin{align}
    & \xi_t \geq 0, & \forall t \in \mathcal{H}^{\rm{train}},\label{eq:10b}\\
    & \xi_t \geq \Delta p_t^{\rm{ABS}} - VaR, & \forall t \in \mathcal{H}^{\rm{train}},\label{eq:10c}
\end{align}
\end{subequations}
with the left hand side of \blueref{eq:10a} being equivalent to the CVaR, constrained by the given parameter $\overline{\Delta P}^{\rm{CVaR}}$. The new auxiliary variables $VaR\in \mathbb{R}$ and $\xi_t\in \mathbb{R}$ must be added to the set of variables $\Omega$. Depending on the underlying operational condition on the permissibility of purchasing power from the grid, three possible trading models arise:
$\mathcal{T}_{\rm{CVaR}}$ (buying is allowed), $\mathcal{T}_{\rm{CVaR}}^{\rm{res}}$ (buying is not allowed), and $\mathcal{T}_{\rm{CVaR}}^{\rm{cond}}$ (buying is conditionally allowed).
\\

\item \textbf{Extreme imbalance ($\mathcal{T}_{\rm{ext}}$, $\mathcal{T}_{\rm{ext}}^{\rm{res}}$, $\mathcal{T}_{\rm{ext}}^{\rm{cond}}$):}\
Here, we enforce a limit on the maximum imbalance $\Delta p_t^{\rm{ABS}}$ such that:
\begin{align}\label{eq:11}
    & \Delta p_t^{\rm{ABS}} \leq \overline{\Delta P}^{\rm{ext}}, & \forall t \in \mathcal{H}^{\rm{train}},
\end{align}
with the parameter $\overline{\Delta P}^{\rm{ext}}$ being the given upper limit on the extreme imbalance. Similarly, we end up in three trading models with extreme imbalance constraints: $\mathcal{T}_{\rm{ext}}$ when buying power from the grid is allowed, $\mathcal{T}_{\rm{ext}}^{\rm{res}}$ when it is not allowed, and $\mathcal{T}_{\rm{ext}}^{\rm{cond}}$ when it is conditionally allowed.
\end{enumerate}

\vspace{1mm}
\subsection{Curve Structuring and Interpretation}\label{sec:bidding curve}
Recall that the feature vector for every hour $t$ includes the realized day-ahead electricity price  as its first element, $x_{t,1}=\hat{\lambda}_t^{\rm{DA}}$. Let us denote $\mathbf{q}_{j,k}^{\rm{DA}}=\{{q}_{j,k,1}^{\rm{DA}},\dots,{q}_{j,k,N}^{\rm{DA}}\}$ and $\mathbf{q}_{j,k}^{\rm{H}}=\{{q}_{j,k,1}^{\rm{H}},\dots,{q}_{j,k,N}^{\rm{H}}\}$. We also define subsets $\mathbf{\Breve{q}}_{j,k}^{\rm{DA}}=\{{{q}}_{j,k,2}^{\rm{DA}},\dots,{{q}}_{j,k,N}^{\rm{DA}}\}$, $\mathbf{\Breve{q}}_{j,k}^{\rm{H}}\{{{q}}_{j,k,2}^{\rm{H}},\dots,{{q}}_{j,k,N}^{\rm{H}}\}$, and $\mathbf{\Breve{x}}_t = \{x_{t,2},\dots,x_{t,N}\}$. This enables the calculation of the day-ahead power bid and the electrolyzer power consumption from \blueref{eq:2} for each hour and price domain as a linear function of the day-ahead electricity price $\lambda^{\rm{DA}}$, i.e., $f(\lambda^{\rm{DA}})$ and $g(\lambda^{\rm{DA}})$, such that:
\begin{subequations}\label{eq:12}
\begin{align} 
p_t^{\rm{DA}} = f(\lambda^{\rm{DA}}) = a_1\lambda^{\rm{DA}}+b_1,\\
p_t^{\rm{H}} = g(\lambda^{\rm{DA}}) = a_2\lambda^{\rm{DA}}+b_2,
\end{align}
\end{subequations}
with the slopes ${a_1 = q_{j,k,1}^{\rm{DA}}}$ and ${a_2 = q_{j,k,1}^{\rm{H}}}$, as well as the intercepts ${b_1\!=\!\mathbf{\Breve{q}}_{j,k}^{\rm{DA}} \ \mathbf{\Breve{x}}_t^\top}$ and ${b_2 = \mathbf{\Breve{q}}_{j,k}^{\rm{H}} \ \mathbf{\Breve{x}}_t^\top}$.

\begin{figure*}[!t]
\centering
\subfloat{\includegraphics[width=2.35in]{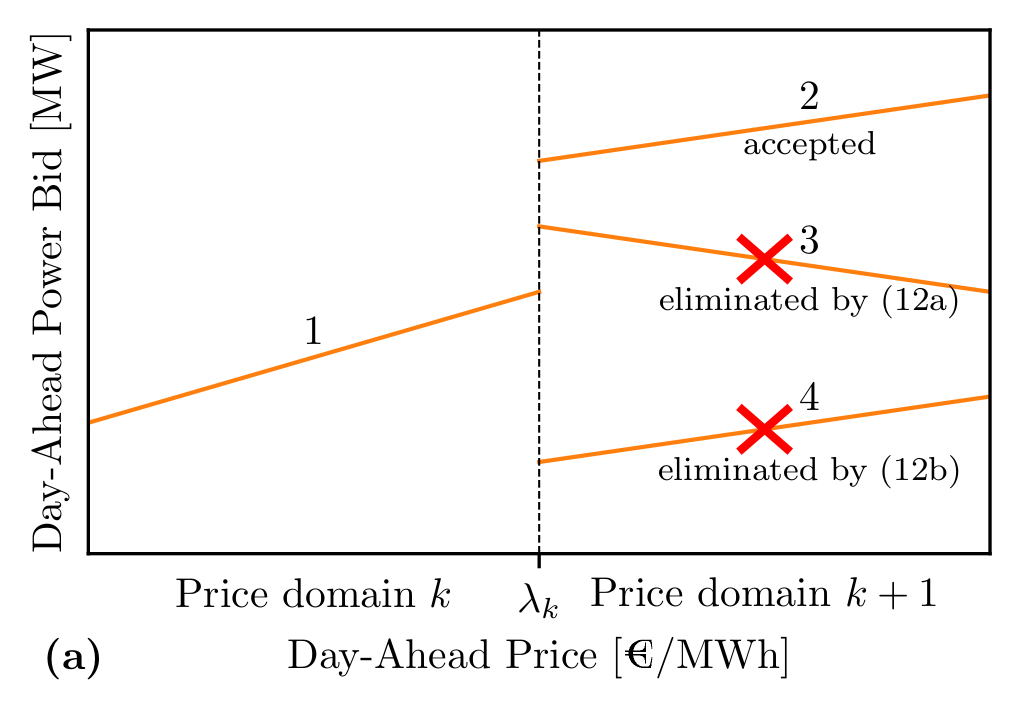}%
\label{fig:bidding_curve_c}}
\hfil
\subfloat{\includegraphics[width=2.35in]{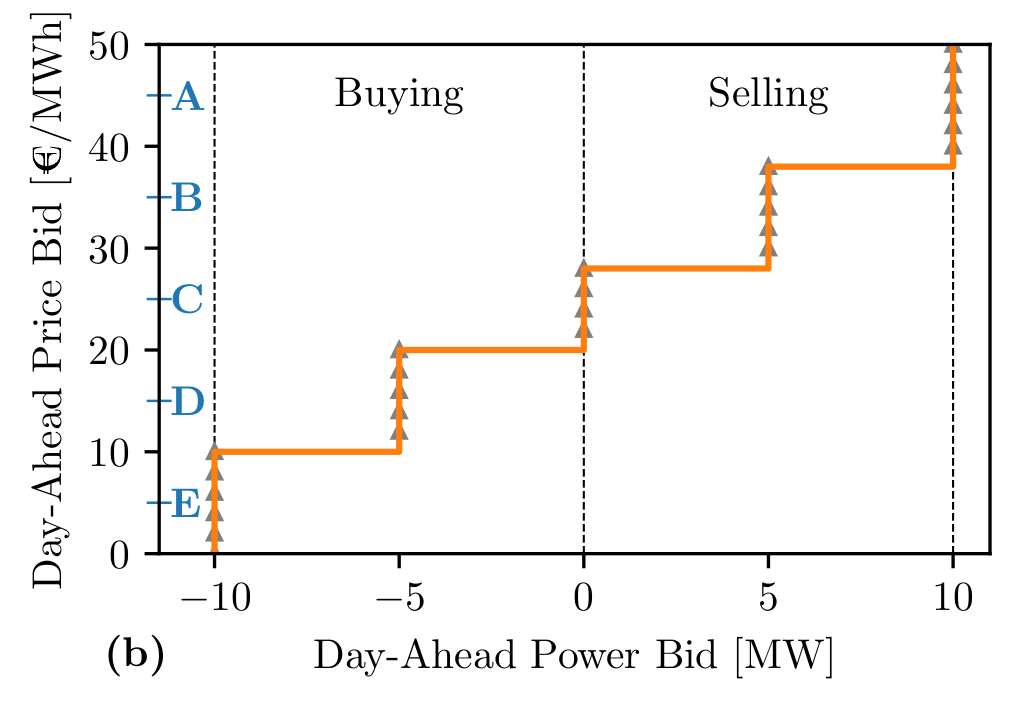}%
\label{fig:bidding_curve_a}}
\hfil
\subfloat{\includegraphics[width=2.35in]{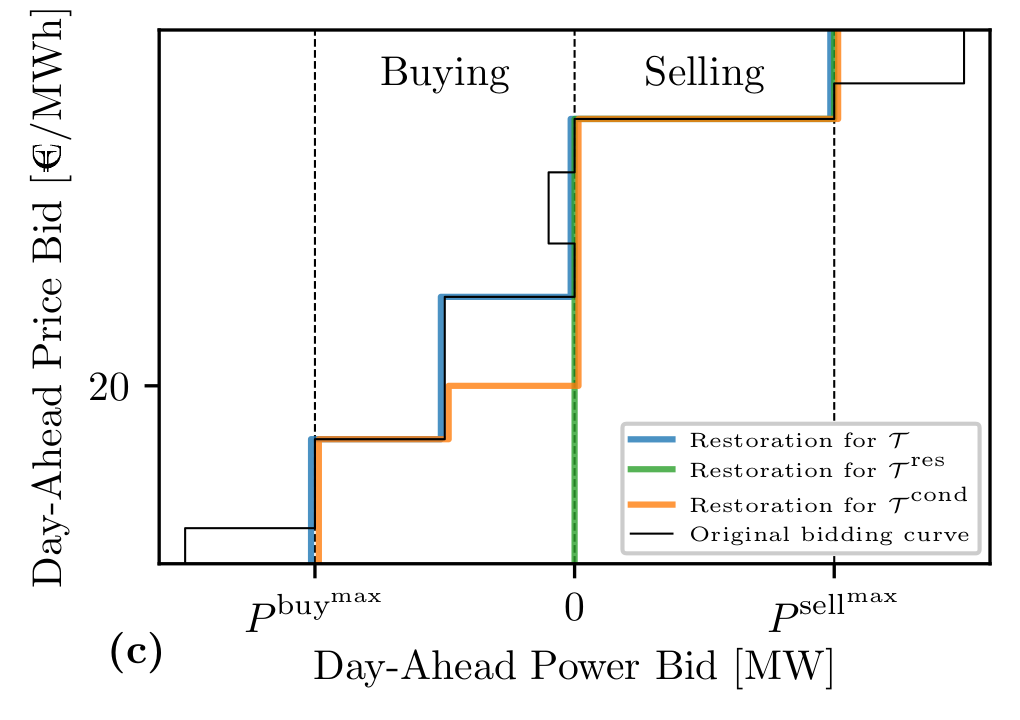}%
\label{fig:bidding_curve_b}}
\caption{Bidding curve examples: 
Plot (a) illustrates the transition from one price domain $k$ to the next price domain $k+1$. Plot (b) shows the discretization (gray triangles) of the bidding function \blueref{eq:12} and the resulting bidding curve (orange). 
Plot (c) visualizes the feasibility restoration of the resulting bidding curves. The original curve (black) is modified to stay within the feasible space for the different models $\mathcal{T}$, $\mathcal{T}^{\rm{res}}$, and $\mathcal{T}^{\rm{cond}}$.}
\label{fig:bidding_curve}
\end{figure*}

To ensure that the resulting curve is non-decreasing in price, as required by most real-world electricity markets, we enforce the following additional constraints:
\begingroup
\allowdisplaybreaks
\begin{subequations}\label{eq:4}
\begin{align}
    & \qquad {q}_{j,k,1}^{\rm{DA}} \geq 0, & \forall k \in \mathcal{K}, \ \forall j \in [1,...,24] \label{eq:4a}\\
    & \qquad ({q}_{j,k+1,1}^{\rm{DA}}-{q}_{j,k,1}^{\rm{DA}})\lambda_k + \nonumber\\
    & \qquad\ \ \ \ \ \ 
    (\mathbf{\Breve{q}}_{j,k+1}^{\rm{DA}}-\mathbf{\Breve{q}}_{j,k}^{\rm{DA}} ) \mathbf{\Breve{x}}_t  \geq 0, &  \forall t \in \mathcal{H}^{\rm{train}}, \ \forall k \in \mathcal{K}, \label{eq:4b}
\end{align}
\end{subequations}
\endgroup
where $\lambda_k$ is the price threshold between price domains $k$ and $k+1$. For example, in the case of two price domains, (0, 100) €/MWh and (100, 1000) €/MWh,  $\lambda_k$=100. 
Constraint \blueref{eq:4a} enforces the resulting bidding curve to be non-decreasing
with respect to the price in each specific price domain $k$. Additionally, \blueref{eq:4b} ensures that the curve remains non-decreasing from price domain $k$ to the next one, $k+1$. In other words, it guarantees that the corresponding quantity bid in the next price domain will be at the same level or a higher level than in the current price domain.
Examples illustrating how these two constraints function are provided in \figref{fig:bidding_curve_c}. Curve 1 represents a sample bidding curve within price domain $k$,  while curves 2, 3, and 4 depict possible bidding curves for domain $k+1$. Constraint \blueref{eq:4a} prevents the formation of curve 3, which has a negative slope, while \blueref{eq:4b} prohibits curve 4, where the bidding curve in price domain $k+1$ is lower than in price domain $k$.

Adding the curve formation constraints \blueref{eq:4} and the different risk constraints added to our three betting models \{$\mathcal{B}$, $\mathcal{B}^{\rm{res}}$, $\mathcal{B}^{\rm{cond}}$\}, we end up in a set of 9  trading models \{$\mathcal{T}_{\rm{mean}}$, $\mathcal{T}_{\rm{CVaR}}$, $\mathcal{T}_{\rm{ext}}$, $\mathcal{T}_{\rm{mean}}^{\rm{res}}$, $\mathcal{T}_{\rm{CVaR}}^{\rm{res}}$, $\mathcal{T}_{\rm{ext}}^{\rm{res}}$, $\mathcal{T}_{\rm{mean}}^{\rm{cond}}$, $\mathcal{T}_{\rm{CVaR}}^{\rm{cond}}$, $\mathcal{T}_{\rm{ext}}^{\rm{cond}}$\}. A full overview of all proposed betting and trading models can be found in \tabref{Tab:model_overview}.

\vspace{1mm}
\section{Testing and Feasibility Restoration}\label{Sec:Test+Feasibility}

\begin{table}[!t]
\renewcommand{\arraystretch}{1.2}
\caption{An overview of all proposed models}
\label{Tab:model_overview}
\centering
\renewcommand{\arraystretch}{1.3} 
\begin{tabular}{lccc}
\hline
 & \scriptsize{Buying} & \scriptsize{Buying not} & \scriptsize{Buying conditionally}\\
& \scriptsize{permitted} & \scriptsize{permitted} & \scriptsize{permitted} \\
&(.) &  (.)$^{\rm{res}}$&  (.)$^{\rm{cond}}$\\
\hline
$\mathcal{B}$ & \blueref{eq:3},\blueref{eq:4},\blueref{eq:5} & \blueref{eq:3},\blueref{eq:4},\blueref{eq:6} & \blueref{eq:3},\blueref{eq:4},\blueref{eq:7}\\
 \hline
$\mathcal{T}_{\rm{mean}}$ & \blueref{eq:3},\blueref{eq:4},\blueref{eq:5}& \blueref{eq:3},\blueref{eq:4},\blueref{eq:6} & \blueref{eq:3},\blueref{eq:4},\blueref{eq:7}\\
& \blueref{eq:8},\blueref{eq:9} & \blueref{eq:8},\blueref{eq:9} & \blueref{eq:8},\blueref{eq:9}\\
\hline
$\mathcal{T}_{\rm{CVaR}}$ & \blueref{eq:3},\blueref{eq:4},\blueref{eq:5} & \blueref{eq:3},\blueref{eq:4},\blueref{eq:6} & \blueref{eq:3},\blueref{eq:4},\blueref{eq:7}\\
& \blueref{eq:8},\blueref{eq:10} & \blueref{eq:8},\blueref{eq:10} & \blueref{eq:8},\blueref{eq:10}\\
\hline
$\mathcal{T}_{\rm{ext}}$ & \blueref{eq:3},\blueref{eq:4},\blueref{eq:5} & \blueref{eq:3},\blueref{eq:4},\blueref{eq:6} & \blueref{eq:3},\blueref{eq:4},\blueref{eq:7}\\
& \blueref{eq:8},\blueref{eq:11} & \blueref{eq:8},\blueref{eq:11} & \blueref{eq:8},\blueref{eq:11}\\
\hline
\end{tabular}
\end{table}
\renewcommand{\arraystretch}{1} 

\subsection{Testing Step}
We now discuss the testing phase, which corresponds to Step (2) in \figref{fig:framework}. Recall that the linear decision policies $\mathbf{q}_{j,k}^{\rm{DA}}$ and $\mathbf{q}_{j,k}^{\rm{H}}$ are fixed to the values obtained in the training phase. Starting with \blueref{eq:12}, the linear bidding curves are derived based on the feature vector $\Breve{\mathbf{x}}_\tau$, which excludes realized prices.
Remember that the first elements of the policies, $q_{j,k,1}^{\rm{DA}}$ and $q_{j,k,1}^{\rm{H}}$, determine the slope of the bidding curves, while the remaining elements, $\Breve{\mathbf{q}}_{j,k}^{\rm{DA}}$ and $\Breve{\mathbf{q}}_{j,k}^{\rm{H}}$, are used to calculate the intercepts. This results in the formation of power trades $p_\tau^{\rm{DA}}$ and electrolyzer consumption schedules $p_\tau^{\rm{H}}$ during the training period, both as functions of the day-ahead electricity price $\lambda^{\rm{DA}}$. Discretizing these functions with respect to $\lambda^{\rm{DA}}$ generates piecewise linear price-quantity curves for both the power bid and the electrolyzer consumption schedule.

These curves are plotted with the day-ahead price $\lambda^{\rm{DA}}$ on the y-axis and the power bid or the electrolyzer consumption on the x-axis, following the standard format used in energy markets, such as Nord Pool. An example of the resulting bidding curve for day-ahead power bids is shown in \figref{fig:bidding_curve_a}. Similarly, a bidding curve can be plotted for the electrolyzer's power consumption. Negative power bids correspond to buying quantities, while positive bids represent selling quantities. 

The final quantity of electricity bought or sold is determined by the realized day-ahead electricity price $\hat{\lambda}_{\tau}^{\rm{DA}}$. To illustrate this, let us refer to \figref{fig:bidding_curve_a}, which shows five price examples for $\hat{\lambda}_{\tau}^{\rm{DA}}$, labeled A to E.  If $\hat{\lambda}_{\tau}^{\rm{DA}} = \rm{A}$, , the HPP operator will sell power to the grid at full capacity, 10 MW, in the day-ahead market. If $\hat{\lambda}_{\tau}^{\rm{DA}}= \rm{B}$, the operator will sell only part of its capacity, specifically 5 MW, which is less than the full 10 MW. If $\hat{\lambda}_{\tau}^{\rm{DA}} = \rm{C}$, the operator will neither sell nor buy power. If  $\hat{\lambda}_{\tau}^{\rm{DA}} = \rm{D}$, the operator will buy 5 MW of power from the grid, and if $\hat{\lambda}_{\tau}^{\rm{DA}} = \rm{E}$, the operator will buy power at full capacity, 10 MW. In all cases, the HPP operator will settle any potential imbalances $\Delta p_\tau$ in the balancing market, based on the realized wind power $\hat{P}_\tau^{\rm{W}}$ and the realized single imbalance price $\hat{\lambda}_{\tau}^{\rm{B}}$. This allows the calculation of the final profit for each training hour $\tau$.

\vspace{1mm}
\subsection{Feasibility Restoration Step} \label{sec: Feasibility Restoration Step}
The model outputs $p_\tau^{\rm{DA}}$ and $p_\tau^{\rm{H}}$ in the testing phase are not necessarily feasible given the  feature vector $\mathbf{\breve{x}}_\tau$. To ensure a feasible solution, we consider the feasibility restoration step, referred to as Step (3) in \figref{fig:framework}. The objective is to find the closest feasible solution if a specific constraint is violated. In practice, this is achieved by modifying the resulting bidding curves and projecting them onto the nearest feasible curve, as illustrated in \figref{fig:bidding_curve_b}.
Assume that the black bidding curve represents the output from the training stage. The first adjustment is to correct the bidding curve to ensure it is always increasing, thereby guaranteeing compliance with \blueref{eq:4}. Based on our numerical results, we have observed that violations of this constraint are rare. For the trading models $\mathcal{T} =$ $\{\mathcal{T}_{\rm{mean}}$, $\mathcal{T}_{\rm{CVaR}}$, $\mathcal{T}_{\rm{ext}}\}$, we further adjust the bidding curve to ensure it stays within the maximum buying and selling quantities (blue). For the trading models $\mathcal{T}^{\rm{res}}=$ $\{\mathcal{T}_{\rm{mean}}^{\rm{res}}$, $\mathcal{T}_{\rm{CVaR}}^{\rm{res}}$, $\mathcal{T}_{\rm{ext}}^{\rm{res}}\}$, where power purchase from the grid is not allowed, we remove all negative power bids, setting them to 0 MW (green).  For the trading model $\mathcal{T}^{\rm{cond}}=$ $\{\mathcal{T}_{\rm{mean}}^{\rm{cond}}$, $\mathcal{T}_{\rm{CVaR}}^{\rm{cond}}$, $\mathcal{T}_{\rm{ext}}^{\rm{cond}}\}$, which includes a price condition for power purchase permission, negative power bids with a price of €20/MWh or more are adjusted to 0 MW (orange). Similarly, feasibility restoration can be applied to the electrolyzer's power consumption, ensuring that the output remains within the feasible operating range.

\begin{figure*}[!t]
    \centering
    \subfloat{
        \includegraphics[width=2.5in]{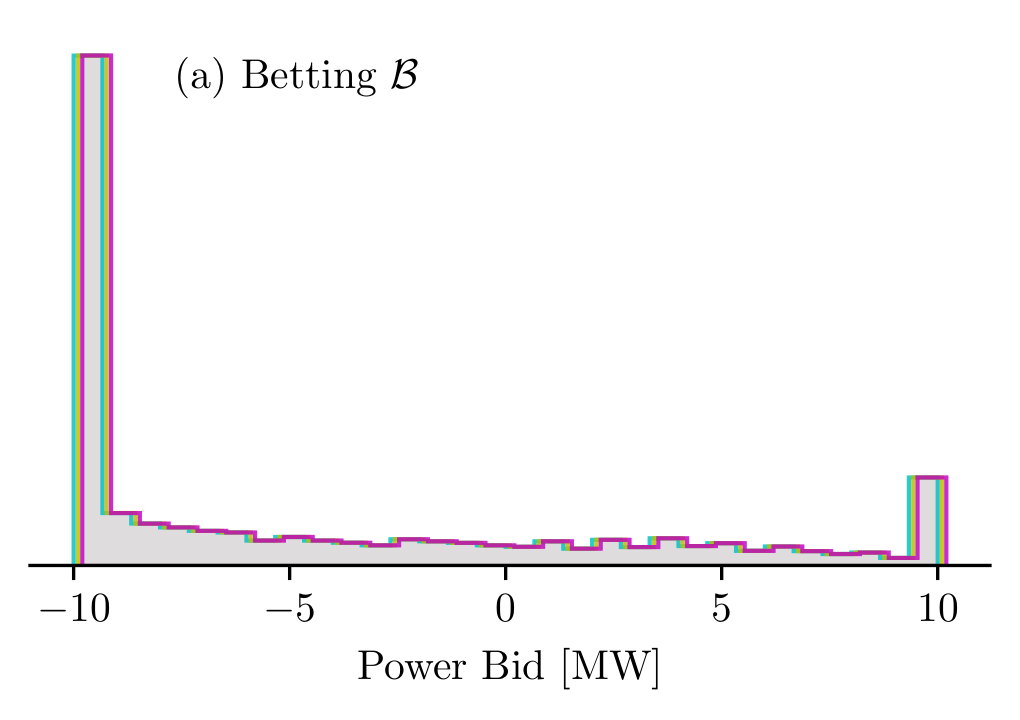}
        \label{fig:distributions_a}
    }
    \hfil
    \subfloat{
        \includegraphics[width=2.5in]{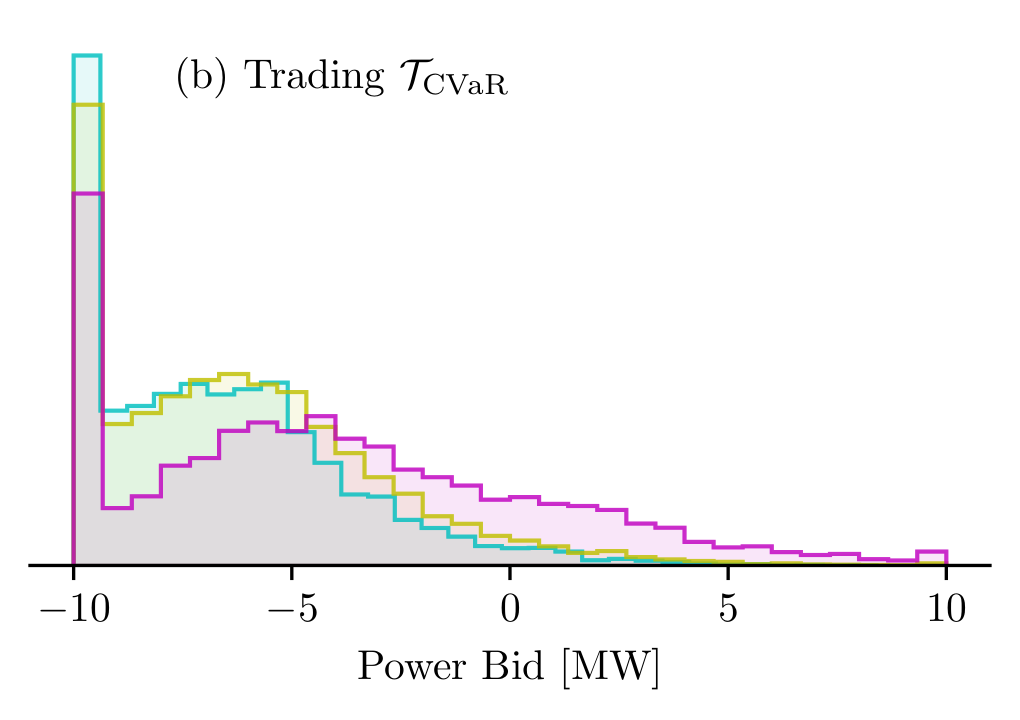}
        \label{fig:distributions_b}
    }
    \\
    \vspace{-1cm}
    \subfloat{
        \raisebox{2cm}{\includegraphics[trim={3cm 2cm 3cm 0.8cm},clip,width=0.75in]{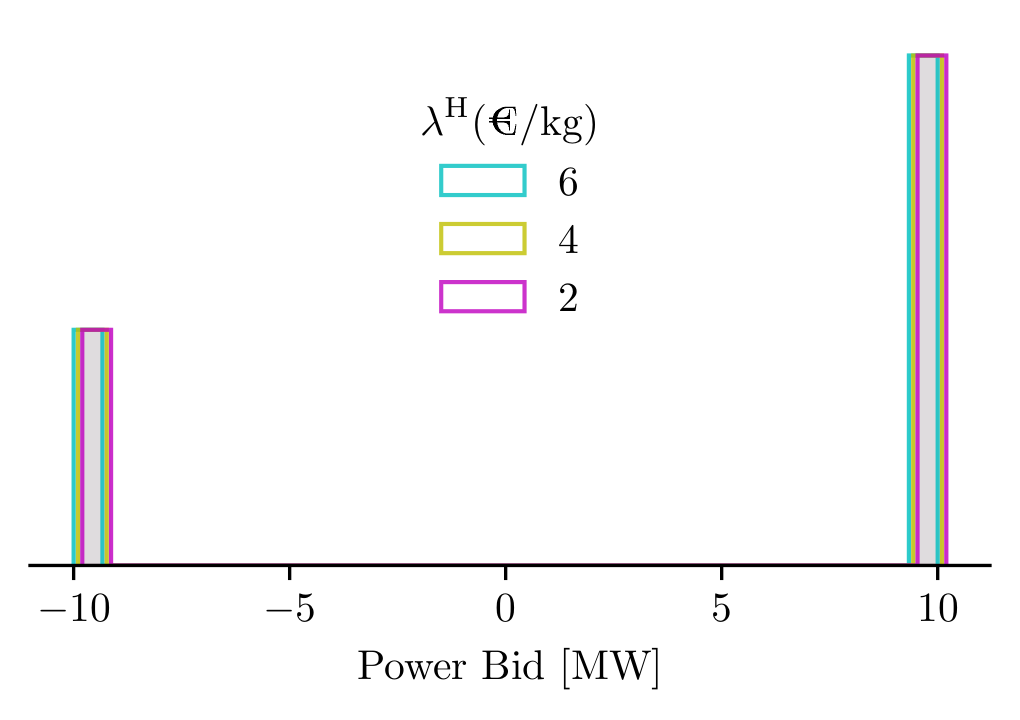}}
    }
    \\
    \vspace{-3.8cm}
    \subfloat{
        \includegraphics[width=2.5in]{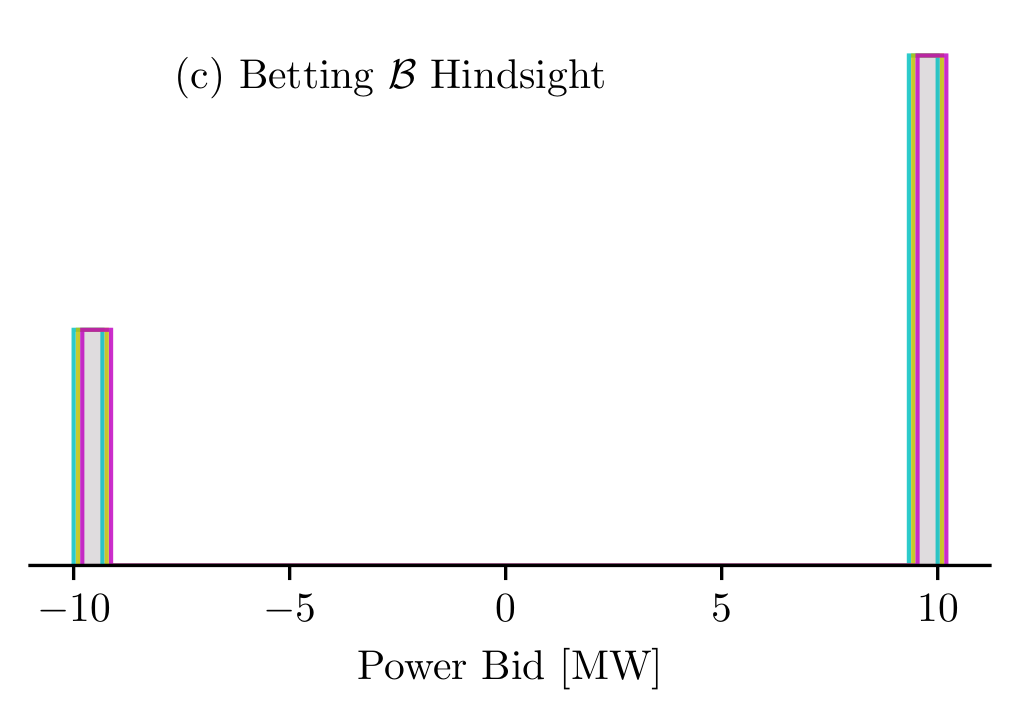}
        \label{fig:distributions_c}
    }
    \hfil
    \subfloat{
        \includegraphics[width=2.5in]{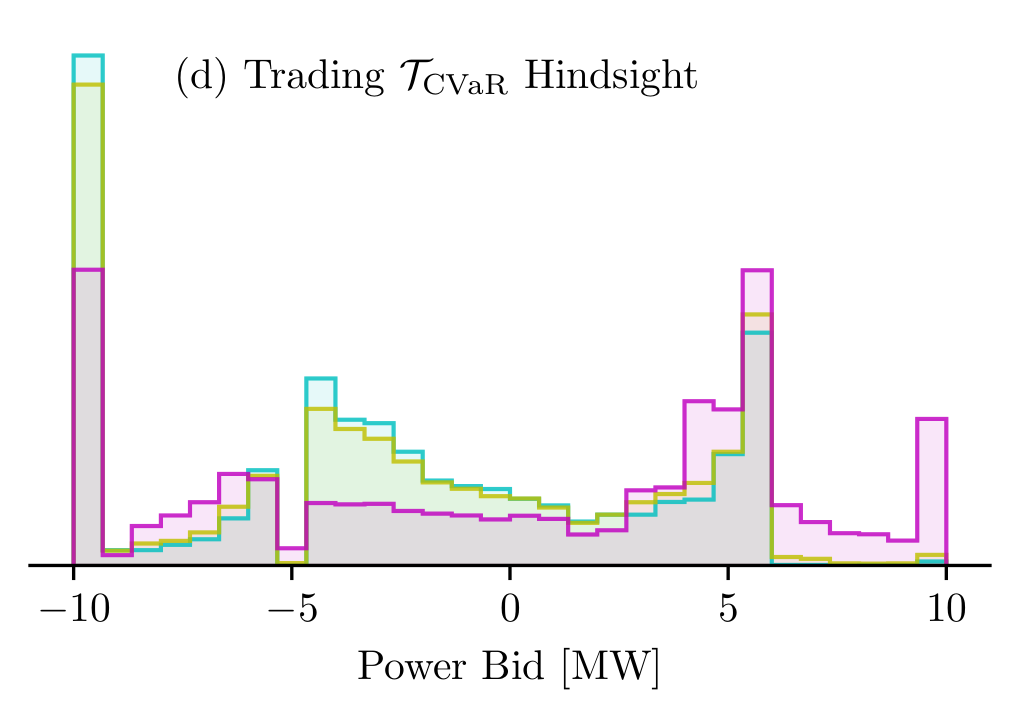}
        \label{fig:distributions_d}
    }
    \caption{The distribution of realized hourly trades $\hat{p}_\tau^{\rm{DA}}$ made by the HPP operator under models $\mathcal{B}$ (plot a) and $\mathcal{T}_{\rm{CVaR}}$ (plot b) is shown for three hydrogen prices: €2/kg, €4/kg, and €6/kg. Plots (a) and (b) display the results obtained from the proposed learning method, while plots (c) and (d) show the results in hindsight, assuming perfect foresight into future realizations.} 
    \label{fig:Distributions
    }
\end{figure*}

\vspace{1mm}
\section{Numerical Results}\label{sec:results}
This section presents a numerical analysis of the proposed framework. All source codes are publicly available in \cite{betting_vs_trading}.

For our study, we use synthetic data generated by \cite{HELGREN2024}, which is based on historical data from the wind farm in Roedsand, Denmark. The dataset includes forecasted and realized wind power production, forecasted and realized day-ahead prices, and aggregated wind forecasts. Additionally, we incorporate single imbalance prices obtained from Energinet. The features used in this case study include the forecasted wind power production and day-ahead price and aggregated wind forecasts. Additionally, the realized day-ahead electricity price is added together with a constant feature 1 as explained in Section \ref{sec: feature set}. For training, we use historical data from the year 2019, and for testing, we use data from 2020. In this case study, the model is trained once on the entire training dataset; however, it is also possible to retrain the model periodically, for example, on a daily basis. 

Following \cite{HELGREN2024} and \cite{RAHELI2023}, the wind farm has a capacity $\overline{P}^{\rm{W}}$ of 10 MW, which is also considered to be the capacity of the electrolyzer $\overline{P}^{\rm{H}}$. The minimum consumption level of the electrolyzer, $\underline{P}^{\rm{H}}$, is set at 15\% of its capacity. The electrolyzer’s production is modeled with two linear segments ($|\mathcal{S}|=2$), where the first segment connects production at minimum power consumption to the production at maximum efficiency, and the second segment connects maximum efficiency to production at maximum power consumption. The minimum hydrogen production requirement, $\underline{D}$, is set to 880 kg per day, which corresponds to a daily electrolyzer consumption of approximately 5 MWh.  The constant efficiency factor for compressing and storing the hydrogen is set to 0.88. For the linear policies, we choose 10 price domains ($|\mathcal{K}|=10$), with each boundary placed at the 10\% quantiles of the historic day-ahead prices.

\vspace{1mm}
\subsection{Betting vs. Trading}
We illustrate how the day-ahead trading decisions of the HPP operator become more diversified when transitioning from a betting to a trading model. We consider an operational condition where the HPP operator is allowed to buy from the grid without any limit. In the trading model, we specifically focus on the case where the imbalance magnitude is constrained by CVaR constraints. Thus, we compare two models: $\mathcal{B}$ (the betting model) and $\mathcal{T}_{\rm{CVaR}}$ (the trading model with CVaR constraints). The parameter $\overline{\Delta P}^{\rm{CVaR}}$ in \blueref{eq:10a} is set to 30\% of the CVaR value obtained from the optimal solution of the betting model, where no explicit constraint limits the magnitude of the imbalance.

\figref{fig:distributions_a} and \figref{fig:distributions_b} show the distribution of realized hourly trades $\hat{p}_\tau^{\rm{DA}}$ made by the HPP operator under models $\mathcal{B}$ and $\mathcal{T}_{\rm{CVaR}}$, respectively, during the training period, which covers all hours of 2020. Both models were trained using data from 2019, which was included in the feature vector $\mathbf{x}_t$ and the realization vector $\mathbf{y}_t$. The linear decision policies were derived from this data and treated as fixed parameters during the testing phase. In the testing phase, the feature vector $\mathbf{\breve{x}}_\tau$ is provided based on 2020 data, feasibility restoration is applied if necessary, and the realized trades are determined based on the realized day-ahead prices from 2020. These plots display three distinct distributions corresponding to three different hydrogen prices ($\lambda^{\rm{H}}$): €2/kg, €4/kg, and €6/kg.

The realized day-ahead power trades in model $\mathcal{T}_{\rm{CVaR}}$ are more diversified than in model $\mathcal{B}$. This diversification can be further adjusted by tuning $\overline{\Delta P}^{\rm{CVaR}}$. In model $\mathcal{B}$, the majority of realized trades are 10 MW, either buying or selling, resulting in a binary decision in many hours. In model $\mathcal{T}_{\rm{CVaR}}$, a higher hydrogen price (€6/kg) leads to more power being purchased in the day-ahead stage, as it becomes more profitable to produce additional hydrogen. Conversely, when the hydrogen price is lower (€2/kg), the HPP operator adopts a more selling-oriented position in the day-ahead electricity market. It is worth noting that the distribution shift between €2/kg and €4/kg is more significant than between €4/kg and €6/kg, suggesting a critical hydrogen price in the €2–€4 range. This price threshold incentivizes maximum hydrogen production, making it more profitable. As a result, there are barely any selling bids from model $\mathcal{T}_{\rm{CVaR}}$ for hydrogen prices of €4/kg or €6/kg.

To verify their ex-post performance, we  present the trades in \textit{hindsight} for both models, $\mathcal{B}$ and $\mathcal{T}_{\rm{CVaR}}$, as illustrated in \figref{fig:distributions_c} and \figref{fig:distributions_d}, respectively. In hindsight, the models hypothetically have access to the realizations $\mathbf{y}_\tau$ at the time of day-ahead decision-making, allowing them to make optimal decisions without the need for training or linear decision policies. According to \figref{fig:distributions_c}, the trading decisions in the betting model are fully binary in nature, whereas this is not the case in \figref{fig:distributions_d}. The trading decisions shown in plots (a) and (b) differ to some extent from those in hindsight in plots (c) and (d), illustrating the potential \textit{regret} the HPP operator may experience and highlighting the potential opportunity to further improve the proposed learning models in future work.

\begin{figure}[!t]
\centering
\includegraphics[width=2.8in]{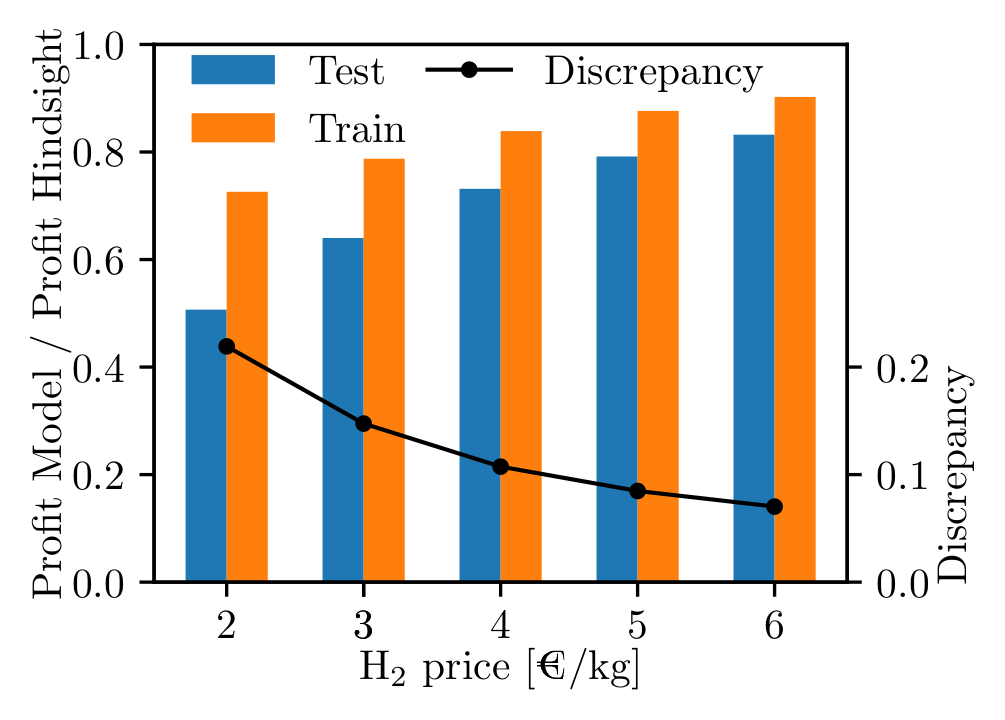}
\caption{Impact of hydrogen price on the profit obtained in both the training and testing phases, as a percentage of the profit in hindsight, for the trading model $\mathcal{T}_{\rm{CVaR}}$.}
\label{fig:h2_prices}
\end{figure}

\vspace{1mm}
\subsection{Impact of the Hydrogen Price}\label{sec:Hydrogen Price}
In the following, we discuss the impact of hydrogen price on: (\textit{i}) the profit discrepancy between hindsight and the training/testing phases, and (\textit{ii}) the profit discrepancy between the training and testing phases. We focus solely on the trading model $\mathcal{T}_{\rm{CVaR}}$. Recall that the profit in the training phase is the optimal value obtained from \blueref{eq:3a}, the profit in the testing phase is the value calculated after feasibility restoration and observing realizations, and the profit in hindsight is the maximum profit that the HPP operator would earn if it had perfect insight into future realizations at the time of day-ahead decision-making. We vary the hydrogen price from €2/kg to €6/kg.

\figref{fig:h2_prices} illustrates both profit discrepancies discussed above. The blue and orange bars represent the \textit{profit ratio} (left y-axis), which is the profit in the training/testing phase divided by the profit in hindsight. Additionally, the difference between the blue and orange bars is indicated by the black curve (right y-axis). As expected, we observe a higher ratio in the training phase compared to the testing phase, as the training model did not have access to the testing data.

The profit ratio  in the testing period increases from 0.5 at a hydrogen price of €2/kg to 0.83 (a 66\% increase) at €6/kg. At the same time, the black curve decreases from 0.22 to 0.07 (a 68\% decrease). This trend can be explained by more straightforward decision-making when the hydrogen price is relatively high. In such cases, the electrolyzer operates at full capacity, consuming local wind power, and, if wind power is insufficient, power purchased from the grid. As a result, the share of profit from hydrogen sales increases, while the profit share from selling wind power in the day-ahead or balancing markets decreases. Conversely, when the hydrogen price is comparatively low, the HPP operator must find an optimal trade-off between hydrogen and wind power sales, which complicates decision-making.

\begin{figure}[!t]
\centering
\includegraphics[width=2.83in]{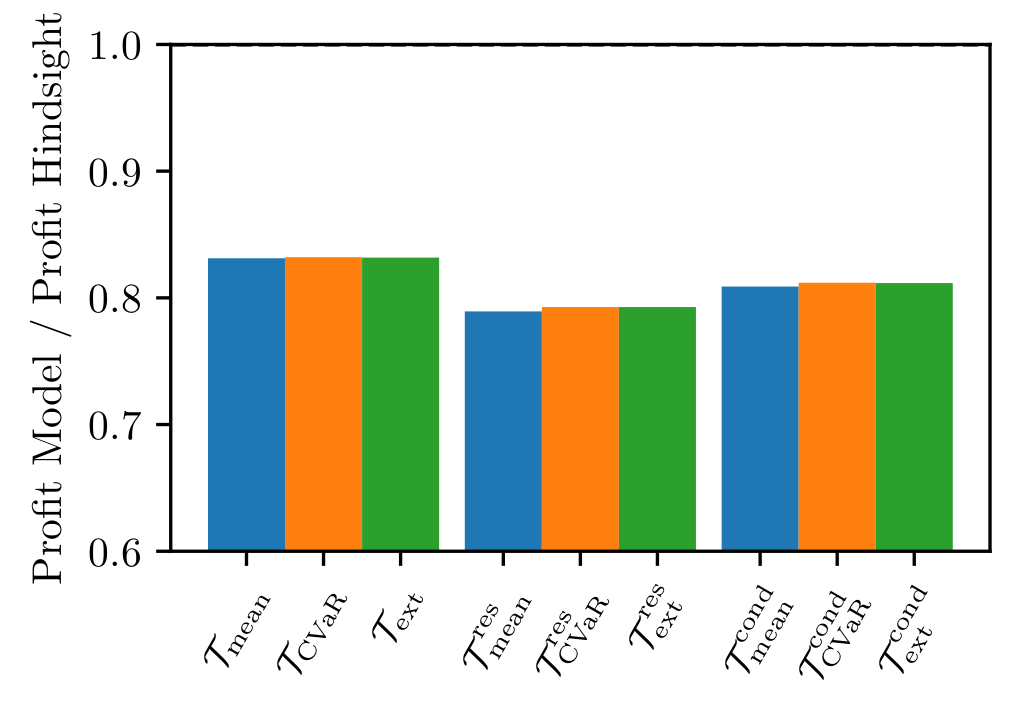}
\caption{Comparison of all nine trading models in terms of their profit during the testing period divided by their profit in hindsight.}
\label{fig:model_comparison}
\end{figure}

\vspace{1mm}
\subsection{Comparison of Nine Trading Models}
Finally, we compare the performance of all nine trading models \{$\mathcal{T}_{\rm{mean}}$, $\mathcal{T}_{\rm{CVaR}}$, $\mathcal{T}_{\rm{ext}}$, $\mathcal{T}_{\rm{mean}}^{\rm{res}}$, $\mathcal{T}_{\rm{CVaR}}^{\rm{res}}$, $\mathcal{T}_{\rm{ext}}^{\rm{res}}$, $\mathcal{T}_{\rm{mean}}^{\rm{cond}}$, $\mathcal{T}_{\rm{CVaR}}^{\rm{cond}}$, $\mathcal{T}_{\rm{ext}}^{\rm{cond}}$\}. Recall that they are different in terms of their permission to buy power from the grid and how the imbalance magnitude is restricted. As in the previous subsection, we use the profit ratio as our benchmark for comparison. 

To ensure a fair comparison, we set the binding limits for all three types of risk constraints, i.e., $\overline{\Delta P}^{\rm{mean}}$, $\overline{\Delta P}^{\rm{CVaR}}$, and $\overline{\Delta P}^{\rm{ext}}$,  to 50\% of their corresponding values in the unconstrained betting model, $\mathcal{B}$. This means, for instance, that in model $\mathcal{T}_{\rm{mean}}$, $\overline{\Delta P}^{\rm{mean}}$ is set to 50\% of the expected value of imbalance settled in model $\mathcal{B}$. The hydrogen price is fixed at €6/kg.


\figref{fig:model_comparison} presents the numerical results for the comparison. The first observation is that the three types of risk constraints show only minor differences in performance, with all discrepancies in the profit ratio being smaller than 0.5\%. In this context, one may conclude that it is sufficient to choose any one of these risk constraints, as they yield nearly equivalent results. However, generalizing this conclusion would require further investigation across different case studies. The second observation is that the best performance, in terms of profit ratio, close to 0.83, is achieved for the unrestricted buying and selling models $\mathcal{T}_{\rm{mean}}$, $\mathcal{T}_{\rm{CVaR}}$, and $\mathcal{T}_{\rm{ext}}$. When the models are restricted to only selling power ( $\mathcal{T}_{\rm{mean}}^{\rm{res}}$, $\mathcal{T}_{\rm{CVaR}}^{\rm{res}}$, $\mathcal{T}_{\rm{ext}}^{\rm{res}}$), their profit ratio drops to around 0.79. For the conditional buying models ($\mathcal{T}_{\rm{mean}}^{\rm{cond}}$, $\mathcal{T}_{\rm{CVaR}}^{\rm{cond}}$, $\mathcal{T}_{\rm{ext}}^{\rm{cond}}$), the profit ratio improves slightly to approximately 0.81. Hence, the lowest performance is observed for the most restricted model.


\vspace{1mm}
\section{Conclusion}\label{sec:conclusion}
In this paper, we develop bidding strategies for day-ahead trading and hydrogen scheduling for a hybrid power plant, which avoid the risky all-or-nothing decisions under single imbalance pricing scheme  by enforcing explicit risk constraints to limit the imbalance magnitude. We show how this approach encourages a more diversified distribution of day-ahead trading decisions. To account for uncertainties at the time of decision-making, we employ linear policies as a practical solution, training them to serve as decision-making guidelines. Additionally, we explain how a feasible bidding curve can be derived from these linear policies using contextual information as input.

We evaluate the model under three operating conditions: when the plant is conditionally permitted, always permitted, or not permitted to buy power from the grid—factors that influence the green certification of the hydrogen produced. By comparing the proposed data-driven bidding strategy in hindsight with perfect information, we show that the data-driven trading models with risk constraints deliver robust and satisfactory performance.

A promising direction would be to explore the limitations of linear decision policies and identify ways to further improve model performance, with the goal of closing the gap between the proposed approach and the hindsight model.  Potential improvements could include enhanced feature selection or the integration of non-linear features, while still maintaining a linear model structure as proposed in \cite{Dvorkin}. This would allow to capture non-linear relationships between the features and the target, while keeping the benefits of a linear model structure. Another possible method is to use the kernel trick, which allows to perform the linear regression in a higher-dimensional space without explicitly computing the transformation as described in \cite{overview}. As a result, non-linear relationships can efficiently be captured by the model. Additionally, it would be valuable to explore opportunities for participation in other trading platforms, such as intraday markets. For the electrolyzer modeling, it is important to gain further insights into whether additional physical and technical details should be considered, such as different operational states, degradation and ramping rates.



\vspace{1mm}
\section*{Acknowledgment}
We gratefully acknowledge the Danish Energy Technology Development and Demonstration Programme (EUDP) for supporting this research through the ViPES2X project (Grant number: 640222-496237), and the Innovation Fund Denmark for supporting our work through the PtX Markets project (Grant number: 150-00001B). We would also like to thank Siemens Gamesa Renewable Energy for providing us with historical forecast data. Additionally, we thank Shobhit Singhal (DTU) for providing valuable feedback on the manuscript.



\vspace{3mm}
\bibliographystyle{IEEEtran}
\bibliography{bib}

\end{document}